\documentclass[twocolumn,showpacs,preprintnumbers,amsmath,amssymb]{revtex4}

\usepackage{graphicx}
\usepackage{dcolumn}
\usepackage{bm}

\begin{document}

\title{Effect of initial configuration on network-based recommendation}
\author{Tao Zhou$^{1,2,3}$}
\email{zhutou@ustc.edu}
\author{Luo-Luo Jiang$^2$}
\author{Ri-Qi Su$^{2,3}$}
\author{Yi-Cheng Zhang$^{1,3}$}
\email{yi-cheng.zhang@unifr.ch}
\affiliation{%
$^1$Department of Physics, University of Fribourg, Chemin du Muse
3, CH-1700 Fribourg, Switzerland \\
$^2$Department of Modern Physics and Nonlinear Science Center,
University of Science and Technology of China, Hefei Anhui,
230026, PR China \\
$^3$Information Economy and Internet Research Laboratory, University
of Electronic Science and Technology of China, Chengdu Sichuan,
610054, PR China
}%

\date{\today}

\begin{abstract}
In this paper, based on a weighted object network, we propose a
recommendation algorithm, which is sensitive to the configuration of
initial resource distribution. Even under the simplest case with
binary resource, the current algorithm has remarkably higher
accuracy than the widely applied global ranking method and
collaborative filtering. Furthermore, we introduce a free parameter
$\beta$ to regulate the initial configuration of resource. The
numerical results indicate that decreasing the initial resource
located on popular objects can further improve the algorithmic
accuracy. More significantly, we argue that a better algorithm
should simultaneously have higher accuracy and be more personal.
According to a newly proposed measure about the degree of
personalization, we demonstrate that a degree-dependent initial
configuration can outperform the uniform case for both accuracy and
personalization strength.
\end{abstract}


\maketitle

\emph{Introduction}. --- The exponential growth of the Internet
\cite{Faloutsos1999} and World-Wide-Web \cite{Broder2000} confronts
people with an information overload: they are facing too many data
and sources to be able to find out those most relevant for them.
Thus far, the most promising way to efficiently filter out the
information overload is to provide personal recommendations. That is
to say, using the personal information of a user (i.e., the
historical track of this user's activities) to uncover his habits
and to consider them in the recommendation. For instances,
Amazon.com uses one's purchase history to provide individual
suggestions. If you have bought a textbook on statistical physics,
Amazon may recommend you some other statistical physics books. Based
on the well-developed \emph{Web 2.0} technology, recommendation
systems are frequently used in web-based movie-sharing
(music-sharing, book-sharing, etc.) systems, web-based selling
systems, and so on. Motivated by the significance to the economy and
society, recommendation algorithms are being extensively
investigated in the engineering community \cite{Adomavicius2005}.
Various kinds of algorithms have been proposed, including
correlation-based methods \cite{Konstan1997,Herlocker2004},
content-based methods \cite{Balab97,Pazzani99}, the spectral
analysis \cite{Maslov00}, principle component analysis
\cite{Goldberg2001}, and so on.

Very recently some physical dynamics, including heat conduction
process \cite{Zhang2007a} and mass diffusion
\cite{Zhou2007,Zhang2007b}, have found applications in personal
recommendation. These physical approaches have been demonstrated to
be both highly efficient and of low computational complexity
\cite{Zhang2007a,Zhou2007,Zhang2007b}. In this paper, we introduce a
network-based recommendation algorithm with degree-dependent initial
configuration. Compared with uniform initial configuration, the
prediction accuracy can be remarkably enhanced by using the
degree-dependent configuration. More significantly, besides the
prediction accuracy, we present novel measurements to judge how
personal the recommendation results are. The algorithm providing
more personal recommendations has, in principle, greater ability to
uncover the individual habits. Since mainstream interests are more
easily uncovered, a user may appreciate a system more if it can
recommend the unpopular objects he/she enjoys. Therefore, we argue
that those two kinds of measurements, accuracy and degree of
personalization, are complementary to each other in evaluating a
recommendation algorithm. Numerical simulations show that the
optimal initial configuration subject to accuracy can also generate
more personal recommendations.

\emph{Method}. --- A recommendation system consists of users and
objects, and each user has collected some objects. Denoting the
object-set as $O=\{o_1,o_2,\cdots,o_n\}$ and user-set as
$U=\{u_1,u_2,\cdots,u_m\}$, the recommendation system can be fully
described by an $n\times m$ adjacent matrix $A=\{a_{ij}\}$, where
$a_{ij}=1$ if $o_i$ is collected by $u_j$, and $a_{ij}=0$ otherwise.
A reasonable assumption is that the objects you have collected are
what you like, and a recommendation algorithm aims at predicting
your personal opinions (to what extent you like or hate them) on
those objects you have not yet collected. Mathematically speaking,
for a given user, a recommendation algorithm generates a ranking of
all the objects he/she has not collected before. The top $L$ objects
are recommended to this user, with $L$ the length of the
recommendation list.

\begin{figure}
\scalebox{0.8}[0.8]{\includegraphics{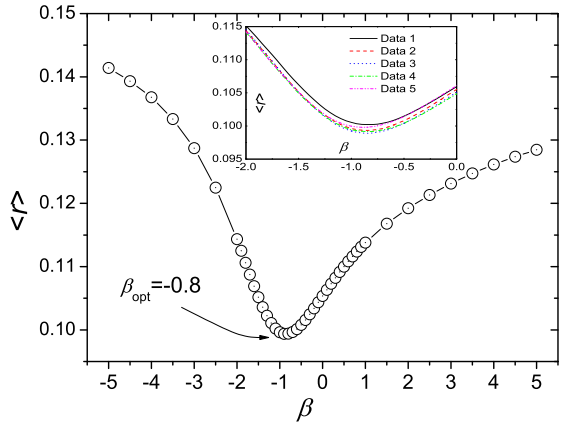}} \caption{(Color
online) The ranking score $\langle r\rangle$ vs. $\beta$. The
optimal $\beta$, corresponding to the minimal $\langle r\rangle
\approx 0.098$, is $\beta_{\texttt{opt}} \approx -0.8$. All the data
points shown in the main plot is obtained by averaging over five
independent runs with different data-set divisions. The inset shows
the numerical results of every separate run, where each curve
represents one random division of data-set.}
\end{figure}

Based on the user-object relations $A$, an object network can be
constructed, where each node represents an object, and two objects
are connected if and only if they have been collected simultaneously
by at least one user. We assume a certain amount of resource (e.g.
recommendation power) is associated with each object, and the weight
$w_{ij}$ represents the proportion of the resource $o_j$ would like
to distribute to $o_i$. For example, in the book-selling system, the
weight $w_{ij}$ contributes to the strength of book $o_i$
recommendation to a customer provided he has bought book $o_j$.
Following a network-based resource-allocation process where each
object distributes its initial resource equally to all the users who
have collected it, and then each user sends back what he/she has
received to all the objects he/she has collected (also equally), the
weight $w_{ij}$ (the fraction of initial resource $o_j$ eventually
gives to $o_i$) can be expressed as:
\begin{equation}
w_{ij}=\frac{1}{k(o_j)}\sum^m_{l=1}\frac{a_{il}a_{jl}}{k(u_l)},
\end{equation}
where $k(o_j)=\sum^n_{i=1}a_{ji}$ and $k(u_l)=\sum^m_{i=1}a_{il}$
denote the \emph{degrees} of object $o_j$ and $u_l$, respectively.
Clearly, the weight between two unconnected objects is zero.
According to the definition of the weighted matrix $W=\{w_{ij}\}$,
if the initial resource vector is $\mathbf{f}$, the final resource
distribution is $\mathbf{f}'=W\mathbf{f}$.

The general framework of the proposed network-based recommendation
is as follows: (i) construct the weighted object network (i.e.
determine the matrix $W$) from the known user-object relations; (ii)
determine the initial resource vector $\mathbf{f}$ for each user;
(iii) get the final resource distribution via
$\mathbf{f}'=W\mathbf{f}$; (iv) recommend those uncollected objects
with highest final resource. Note that the initial configuration
$\mathbf{f}$ is determined by the user's personal information, thus
for different users, the initial configuration is different. From
now on, for a given user $u_i$, we use $\mathbf{f}^i$ to emphasize
this personal configuration.

\emph{Numerical results}. --- For a given user $u_i$, the $j$th
element of $\mathbf{f}^i$ should be zero if $a_{ji}=0$. That is to
say, one should not put any recommendation power (i.e. resource)
onto an uncollected object. The simplest case is to set a uniform
initial configuration as
\begin{equation}
f^i_j=a_{ji}.
\end{equation}
Under this configuration, all the objects collected by $u_i$ have
the same recommendation power. In despite of its simplicity, it can
outperform the two most widely applied recommendation algorithms,
\emph{global ranking method} (GRM) \cite{ex1} and
\emph{collaborative filtering} (CF) \cite{ex2}.

To test the algorithmic accuracy, we use a benchmark data-set,
namely \emph{MovieLens} \cite{ex3}. The data consists of 1682 movies
(objects) and 943 users, and users vote movies using discrete
ratings 1-5. We therefore applied a coarse-graining method similar
to that used in Ref. \cite{Blattner2007}: a movie has been collected
by a user if and only if the giving rating is at least 3 (i.e. the
user at least likes this movie). The original data contains $10^5$
ratings, 85.25\% of which are $\geq 3$, thus after coarse gaining
the data contains 85250 user-object pairs. To test the
recommendation algorithms, the data set is randomly divided into two
parts: The training set contains 90\% of the data, and the remaining
10\% of data constitutes the probe. The training set is treated as
known information, while no information in the probe set is allowed
to be used for prediction.

\begin{figure}
\scalebox{0.8}[0.8]{\includegraphics{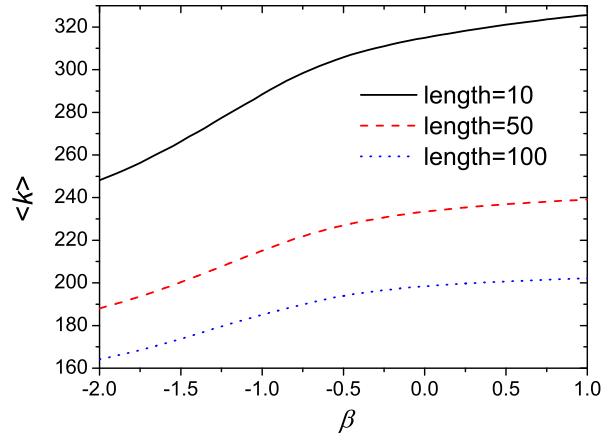}} \caption{(Color
online) The average degree of all recommended movies vs. $\beta$.
The black solid, red dash and blue dot curves represent the cases
with typical lengths $L=10$, 50 and 100, respectively. All the data
points are obtained by averaging over five independent runs with
different data-set divisions. }
\end{figure}

A recommendation algorithm should provide each user with an ordered
queue of all its uncollected objects. For an arbitrary user $u_i$,
if the relation $u_i-o_j$ is in the probe set (according to the
training set, $o_j$ is an uncollected object for $u_i$), we measure
the position of $o_j$ in the ordered queue. For example, if there
are 1000 uncollected movies for $u_i$, and $o_j$ is the 10th from
the top, we say the position of $o_j$ is 10/1000, denoted by
$r_{ij}=0.01$. Since the probe entries are actually collected by
users, a good algorithm is expected to give high recommendations to
them, thus leading to small $r$. Therefore, the mean value of the
position value $\langle r\rangle$ (called \emph{ranking score}
\cite{Zhou2007}), averaged over all the entries in the probe, can be
used to evaluate the algorithmic accuracy: the smaller the ranking
score, the higher the algorithmic accuracy, and vice verse.
Implementing the three algorithms mentioned above, the average
values of ranking scores over five independent runs (one run here
means an independently random division of data set) are 0.107,
0.122, and 0.140 for network-based recommendation, collaborative
filtering, and global ranking method, respectively. Clearly, even
under the simplest initial configuration, subject to the algorithmic
accuracy, the network-based recommendation outperforms the other two
algorithms.

\begin{figure}
\scalebox{0.8}[0.8]{\includegraphics{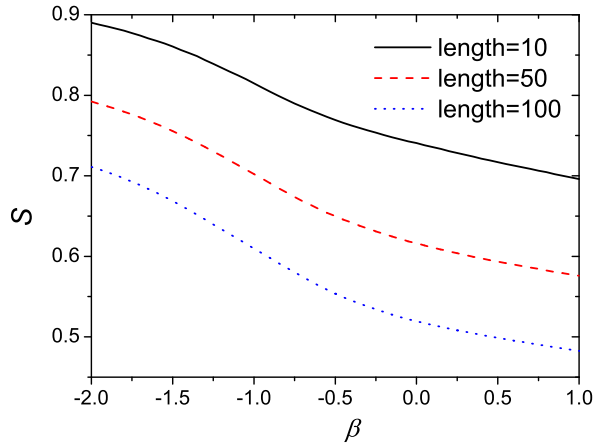}} \caption{(Color
online) $S$ vs. $\beta$. The black solid, red dash and blue dot
curves represent the cases with typical lengths $L=10$, 50 and 100,
respectively. All the data points are obtained by averaging over
five independent runs with different data-set divisions. }
\end{figure}

Consider the initial resource located on object $o_i$ as its
assigned recommendation power. In the whole recommendation process,
the total power given to $o_i$ is $p_i=\sum_jf^j_i$, where the
superscript $j$ runs over all the users $u_j$. Under uniform initial
configuration (see Eq. (2)), the total power of $o_i$ is
$p_i=\sum_jf^j_i=\sum_ja_{ij}=k(o_i)$. That is to say, the total
recommendation power assigned to an object is proportional to its
degree, thus the impact of high-degree objects (e.g. popular movies)
is enhanced. Although it already has a good algorithmic accuracy,
this uniform configuration may be oversimplified, and depressing the
impact of high-degree objects in an appropriate way could, perhaps,
further improve the accuracy. Motivated by this, we propose a more
complicated distribution of initial resource to replace Eq. (2):
\begin{equation}
f^i_j=a_{ji}k^\beta(o_j),
\end{equation}
where $\beta$ is a tunable parameter. Compared with the uniform
case, $\beta=0$, a positive $\beta$ strengthens the influence of
large-degree objects, while a negative $\beta$ weakens the influence
of large-degree objects. In particular, the case $\beta=-1$
corresponds to an identical allocation of recommendation power
($p_i=1$) for each object $o_i$.

Fig. 1 reports the algorithmic accuracy as a function of $\beta$.
The curve has a clear minimum around $\beta=-0.8$. Compared with the
uniform case, the ranking score can be further reduced by 9\% at the
optimal value. It is indeed a great improvement for recommendation
algorithms. Note that $\beta_{\texttt{opt}}$ is close to -1, which
indicates that the more homogeneous distribution of recommendation
power among objects may lead to a more accurate prediction.

Besides accuracy, another significant ingredient one should take
into account to for a personal recommendation algorithm is how
personal this algorithm is. For example, suppose there are 10
perfect movies not yet known for user $u_i$, 8 of which are widely
popular, while the other two fit a certain specific taste of $u_i$.
An algorithm recommending the 8 popular movies is very nice for
$u_i$, but he may feel even better about a recommendation list
containing those two unpopular movies. Since there are countless
channels to obtain information on popular movies (TV, the Internet,
newspapers, radio, etc.), uncovering very specific preference,
corresponding to unpopular objects, is much more significant than
simply picking out what a user likes from the top of the list. To
measure this factor, we go simultaneously in two directions.
Firstly, given the length $L$ of recommendation list, the popularity
can be measured directly by averaging the degree $\langle k\rangle$
over all the recommended objects. One can see from Fig. 2 that the
average degree is positively correlated with $\beta$, thus
depressing the recommendation power of high-degree objects gives
more opportunity to unpopular objects. Also for $L=10$, 50 and 100,
the corresponding $\langle k\rangle$ are 353.50, 258.00 and 214.09
(GRM), as well as 84.62, 87.95 and 83.79 (CF). Since GRM always
recommends the most popular objects, it is clear that $\langle
k\rangle_{\texttt{GRM}}$ is the largest. On the other hand, CF
mainly depends the similarity between users. Thus one user may be
recommended an object collected by another user having very similar
habits to him, even though this object may be very unpopular. This
is the reason why $\langle k\rangle_{\texttt{CF}}$ is the smallest.
Secondly, one can measure the strength of personalization via the
Hamming distance. If the overlapped number of objects in $u_i$ and
$u_j$'s recommendation lists is $Q$, their Hamming distance is
$H_{ij}=1-Q/L$. Generally speaking, a more personal recommendation
list should have larger Hamming distances to other lists.
Accordingly, we use the mean value of Hamming distance $S=\langle
H_{ij}\rangle$, averaged over all the user-user pairs, to measure
the strength of personalization. Fig. 3 plots $S$ vs. $\beta$ and,
in accordance with the numerical results shown in Fig. 2, depressing
the influence of high-degree objects makes the recommendations more
personal. For $L=10$, 50 and 100, the corresponding $S$ are 0.508,
0.397 and 0.337 (GRM), as well as 0.654, 0.501 and 0.421 (CF). Note
that, $S_{\texttt{GRM}}$ is obviously larger than zero, because the
collected objects will not appear in the recommendation list, thus
different users have different recommendation lists. Since CF has
the potential to enhance the user-user similarity, $S_{\texttt{CF}}$
is remarkably smaller than that corresponding to negative $\beta$ in
network-based recommendation.

In a word, without any increase in the algorithmic complexity, using
an appropriate negative $\beta$ in our algorithm outperforms the
uniform case (i.e. $\beta=0$) for all three criteria: more accurate,
less popular, and more personalized.

\emph{Conclusions}. --- In this paper, we propose a recommendation
algorithm based on a weighted object network. This algorithm is
sensitive to the configuration of initial resource distribution.
Even under the simplest case with binary resource, the current
algorithm has remarkably higher accuracy than the widely applied GRM
and CF. Since the computational complexity of this algorithm is much
less than that of CF \cite{ex4}, it has great potential significance
in practice. Furthermore, we introduce a free parameter $\beta$ to
regulate the initial configuration of resource. Numerical results
indicate that decreasing the initial resource located on popular
objects further improves the algorithmic accuracy: In the optimal
case ($\beta_{\texttt{opt}} \approx -0.8$), the distribution of
total initial resource located on each object is very homogeneous
($p_i\sim k^{0.2}(o_i)$). Besides the ranking score, there have been
many measures suggested to evaluate the accuracy of personal
recommendation algorithms
\cite{Zhou2007,Billsus1998,Sarwar2000,Huang2004}, including
\emph{hitting rate}, \emph{precision}, \emph{recall},
\emph{F-measure}, and so on. However, thus far, there has been no
consideration of the degree of personalization. In this paper, we
suggest two measures, $\langle k\rangle$ and $S$, to address this
issue. We argue that to evaluate the performance of a recommendation
algorithm, one should take into account not only the accuracy, but
also the degree of personalization and popularity of recommended
objects. Even under this more strict criterion, the case with
$\beta_{\texttt{opt}} \approx -0.8$ outperforms the uniform case.
Theoretical physics provides us some beautiful and powerful tools in
dealing with this long-standing challenge in modern information
science: how to do a personal recommendation. We believe the current
work can enlighten readers in this interesting direction.

We acknowledge Runran Liu for very valuable discussion and comments
on this work. This work is partially supported by SBF (Switzerland)
for financial support through project C05.0148 (Physics of Risk),
and the Swiss National Science Foundation (205120-113842). TZhou
acknowledges NNSFC under Grant No. 10635040.


\begin{thebibliography} {1}

\bibitem{Faloutsos1999} M. Faloutsos, P. Faloutsos, and C. Faloutsos, Comput. Comm. Rev. {\bf 29}, 251 (1999).
\bibitem{Broder2000} A. Broder, \emph{et al.}, Comput. Netw. {\bf 33}, 309 (2000).
\bibitem{Adomavicius2005} G. Adomavicius, and A. Tuzhilin, IEEE
Trans. Know. \& Data Eng. {\bf 17}, 734 (2005).
\bibitem{Konstan1997} J. A. Konstan, B. N. Miller, D. Maltz, J. L. Herlocker, L. R. Gordon, and J. Riedl, Commun. ACM {\bf 40}, 77 (1997).
\bibitem{Herlocker2004} J. L. Herlocker, J. A. Konstan, K. Terveen, and J. T. Riedl, ACM Trans. Inform. Syst. {\bf 22}, 5 (2004).
\bibitem{Balab97} M. Balabanovi\'c and Y. Shoham, Commun. ACM {\bf 40}, 66 (1997).
\bibitem{Pazzani99} M. J. Pazzani, Artif. Intell. Rev. {\bf 13}, 393 (1999).
\bibitem{Maslov00} S. Maslov, and Y.-C. Zhang,
Phys. Rev. Lett. {\bf 87}, 248701 (2001).
\bibitem{Goldberg2001} K. Goldberg, T. Roeder, D. Gupta, and C.
Perkins, Inf. Ret. {\bf 4}, 133 (2001).
\bibitem{Zhang2007a} Y.-C. Zhang, M. Blattner, and Y.-K. Yu, Phys.
Rev. Lett. {\bf 99}, 154301 (2007).
\bibitem{Zhou2007} T. Zhou, J. Ren, M. Medo, and Y.-C. Zhang, Phys. Rev. E {\bf 76}, 046115 (2007).
\bibitem{Zhang2007b} Y.-C. Zhang, M. Medo, J. Ren, T. Zhou, T. Li, and F. Yang, EPL {\bf 80}, 68003 (2007).
\bibitem{ex1} The global ranking method sorts all the objects in the descending order
of degree and recommends those with highest degrees.
\bibitem{ex2} The collaborative filting is based on measuring the
similarity between users. For two users $u_i$ and $u_j$, their
similarity can be simply determined by
$s_{ij}=\sum^n_{l=1}a_{li}a_{lj} / \texttt{min}\{k(u_i),k(u_j)\}$.
For any user-object pair $u_i-o_j$, if $u_i$ has not yet collected
$o_j$ (i.e., $a_{ji}=0$), the predicted score, $v_{ij}$ (to what
extent $u_i$ likes $o_j$), is given as $v_{ij}=\sum^m_{l=1,l\neq
i}s_{li}a_{jl} / \sum^m_{l=1,l\neq i}s_{li}$. For any user $u_i$,
all the nonzero $v_{ij}$ with $a_{ji}=0$ are sorted in descending
order, and those objects in the top are recommended.
\bibitem{ex3} The MovieLens data can be downloaded from the web-site of
\emph{GroupLens Research} (http://www.grouplens.org).
\bibitem{Blattner2007} M. Blattner, Y. -C. Zhang, and S. Maslov, Physica A {\bf 373}, 753 (2007).
\bibitem{ex4} Instead of calculating all the elements in $W$, one
can implement the current algorithm by directly diffusing the
resource of each user. Ignoring the degree-degree correlation in
user-object relations, the algorithmic complexity is
$\mathbb{O}(m\langle k_u\rangle\langle k_o\rangle)$, where $\langle
k_u\rangle$ and $\langle k_o\rangle$ denote the average degree of
users and objects. Correspondingly, the algorithmic complexity of
collaborative filtering is $\mathbb{O}(m^2\langle
k_u\rangle+mn\langle k_o\rangle)$, where the first term accounts for
the calculation of similarity between users, and the second term
accounts for the calculation of the predictions.
\bibitem{Billsus1998} D. Billsus and M. J. Pazzani, \emph{Proc. 15th Int.
Conf. Machine Learning}, pp. 46-54 (1998).
\bibitem{Sarwar2000} B. Sarwar, G. Karypis, J. Konstan, and J.
Riedl, \emph{Proc. ACM Conf. Electronic Commerce}, pp. 158-167
(2000).
\bibitem{Huang2004} Z. Huang, H. Chen, and D. Zeng, ACM Trans. Inf.
Syst. {\bf 22}, 116 (2004).




\end{thebibliography}
\end{document}